\documentstyle[twoside,fleqn,espcrc2,epsfig,english]{article}
\title{Abelian hard thermal loops on a lattice}

\author{A. Rajantie\address{Centre for Theoretical Physics,
University of Sussex,
Brighton BN1 9QH, United Kingdom}}

\def\lsi{\raise0.3ex\hbox{$<$\kern-0.75em\raise-1.1ex\hbox{$\sim$}}}
\def\gsi{\raise0.3ex\hbox{$>$\kern-0.75em\raise-1.1ex\hbox{$\sim$}}}

\newcommand{\gsim}{\mathop{\gsi}}

\begin{document}
\begin{abstract}
In Abelian theories, one can write 
for the hard thermal loop equations of motions a local formulation that
is more economical than the traditional Vlasov formulation
and is in explicitly canonical form. I show how this formulation can
be used for simulating non-equilibrium
dynamics in the Abelian Higgs model.
\end{abstract}

\maketitle

\section{HARD THERMAL LOOPS}
Understanding the behaviour of quantum fields out of equilibrium is
important both for cosmology and for heavy ion theory.
A particularly interesting example of an event in the early universe
in which out-of-equilibrium dynamics played a crucial role is
the electroweak phase transition, 
since it may explain the observed baryon asymmetry in the 
Universe \cite{ref:baryogenesis}. 
Although static equilibrium properties such as the phase structure 
of the theory have been determined
to high accuracy with numerical lattice simulations~\cite{ref:su2higgs}, 
very little is known
about the non-equilibrium behaviour. The reason is simply that while
static correlators are given by a Euclidean path integral, which
is well-suited for Monte Carlo simulations, one needs to evaluate
a Minkowskian path integral to obtain any real-time correlator.

In the case of the electroweak phase transition, we can safely assume that
the fields are in thermal equilibrium before the transition. The
occupation number of the soft modes (with momentum $k\ll T$) is large, 
and a classical
approximation can be used for them \cite{ref:classical}.
On the other hand, the hard modes ($k\gsim T$) have a small
occupation number and must be treated as quantum fields. 
Perturbation theory works well for the hard modes, and we can
integrate out them
perturbatively, constructing a classical effective theory for the 
soft modes \cite{ref:pointparticles}.
This construction is free of infrared problems, since the loop integrals
have an infrared cutoff, namely the ultraviolet cutoff $\Lambda$ of
the effective theory.
This approach has previously 
been used to measure the sphaleron rate of hot SU(2) gauge 
theory \cite{ref:pointparticles,ref:vlasov}.
It can also be used for simulating non-equilibrium dynamics of
phase transitions,
since the phase of the theory is a property of the soft modes only, 
and the distribution of the hard modes does not change in the
transition. 

At one-loop level, the resulting classical theory is simply the 
hard thermal loop effective theory. In high-temperature approximation
and neglecting the IR cutoff in the loop integrals, we end up
in the expression~\cite{ref:Krammer}
\begin{eqnarray}
{\cal L}_{\rm HTL}&=&-\frac{1}{4}F_{\mu\nu}F^{\mu\nu}\nonumber\\&&
-\frac{1}{4}m_D^2
\int\frac{d\Omega}{4\pi}F^{\mu\alpha}
\frac{v_\alpha v^\beta}{(v\cdot\partial)^2} F_{\mu\beta}\nonumber\\
&&+|D_\mu\phi|^2
-m_T^2|\phi|^2-\lambda|\phi|^4,
\label{equ:lagrHTL}
\end{eqnarray}
where $m_D^2=\frac{1}{3}e^2T^2$, $m_T^2=m^2+(e^2/4+\lambda/3)T^2$, and 
the integration 
is taken over the unit
sphere of velocities $v=(1,\vec{v})$, $\vec{v}^2=1$.
The time evolution of the fields is given by the equations of motion
\begin{eqnarray}
\partial_\mu F^{\mu\nu}&\!=\!&m_D^2\int\!\frac{d\Omega}{4\pi}
\frac{v^\nu v^i}{v\cdot\partial}E^i-2e{\rm Im}
\phi^*D^\nu\phi,
\label{equ:nonlA}
\\
D_\mu D^\mu\phi&\!=\!&-m_T^2\phi-2\lambda(\phi^*\phi)\phi.
\label{equ:nonlf}
\end{eqnarray}
Because of the derivative in the denominator, the gauge field equation of
motion is non-local: the field interacts with fields on its whole 
past light cone.

\section{LOCAL FORMULATION}
To make numerical simulations practical, one needs a local formulation
for the theory. For non-Abelian theories, two such formulations have been
proposed: adding a large number of charged point particles 
\cite{ref:pointparticles} and
introducing an extra field $W$, which satisfies the non-Abelian Vlasov
equation \cite{ref:vlasov,ref:iancu}. 
In practice, both of them lead to a 5+1-dimensional 
field theory.
However, we will use the formulation presented in Ref.~\cite{ref:oma}, 
where it
was shown that in the Abelian case one can integrate
out one of the dimensions, thus obtaining a 4+1-dimensional theory that
is completely equivalent with the others. It consists of two
extra fields $\vec{f}(t,\vec{x},z)$ and $\theta(t,\vec{x},z)$,
where $z\in [0,1]$. In the temporal gauge, they satisfy the equations of
motion
\begin{eqnarray}
\partial_0^2{\vec{f}}(z)&=&z^2\vec{\nabla}^2\vec{f}+m_Dz\sqrt\frac{1-z^2}{2}
\vec\nabla\times\vec{A},\\
\partial_0^2{\theta}(z)&=&z^2\vec{\nabla}\cdot\left(
\vec{\nabla}\theta-m_D\vec{A}\right),\\
\partial_0^2{\vec{A}}&=&-\vec\nabla\times\vec\nabla\times\vec{A}
+m_D\int_0^1dzz^2\left(
\vec\nabla\theta\right.\nonumber\\&&\left.
-m_D\vec{A}+
\sqrt{\frac{1-z^2}{2z^2}}\vec\nabla\times\vec{f}
\right)+\vec{j}.
\end{eqnarray}

It is now possible to calculate any real-time correlator at finite 
temperature. One simply takes a large number of initial configurations
from the thermal ensemble (the explicit form of the Hamiltonian was
given in Ref.~\cite{ref:oma}), and evolves each of them in time, measuring 
the correlator of interest. The average over the initial configurations
gives the ensemble average of the correlator. In principle, that 
approximates the corresponding expectation value in the original quantum
theory to leading order in gauge coupling constant. In practice,
one has to worry about some technical issues concerning the lattice cutoff
that have been discussed in 
Refs.~\cite{ref:pointparticles,ref:vlasov,ref:iancu,ref:oma,ref:bodeker}. 

\section{SIMULATIONS}
To simulate the theory numerically, we discretize it by putting it on a
lattice \cite{ref:oma}. We expand $\vec{f}$ and $\theta$ in terms of Legendre 
polynomials:
\begin{eqnarray}
\vec{f}^{(n)}&=&\int_0^1dzz\sqrt\frac{2}{1-z^2}P_{2n}(z)\vec{f}(z),
\nonumber\\
\theta^{(n)}&=&\int_0^1dzP_{2n}(z)\theta(z),
\label{equ:defLeg}
\end{eqnarray}

In addition to equilibrium real-time correlators, this formalism can be used
to simulate out-of-equilibrium phenomena, such as the dynamics of
the first-order phase transition \cite{ref:u1phases} at $\lambda\ll e^2$.
We do this in three steps:
\begin{enumerate}
\item We thermalize the soft modes to $T>T_c$ with a Monte Carlo simulation.
\item We decrease the temperature to slightly below $T_c$, and perform
a smaller number of sweeps and generate the hard modes around the
soft background. Since we use only local updates, the system
does not tunnel to the true minimum, i.e., the Higgs phase, but remains in
the Coulomb phase.
\item We solve numerically the equations of motion
using the configuration obtained in step 2
as an initial configuration.
\end{enumerate}

\begin{figure}
\epsfig{file=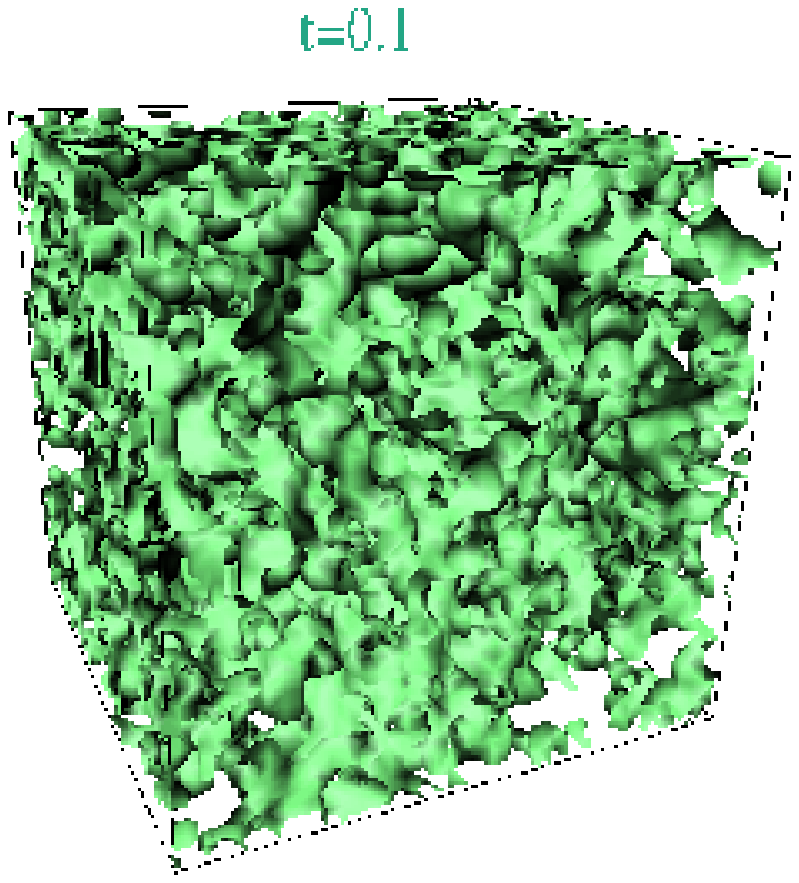,width=5cm}
\vspace{-.4cm}

\epsfig{file=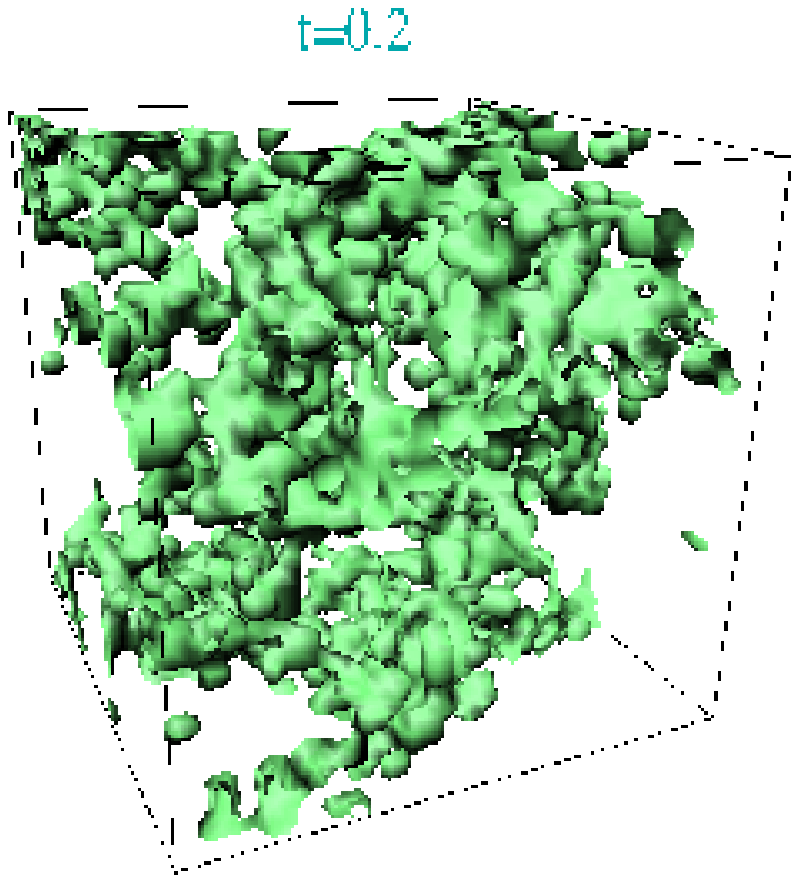,width=5cm}
\vspace{-.4cm}

\epsfig{file=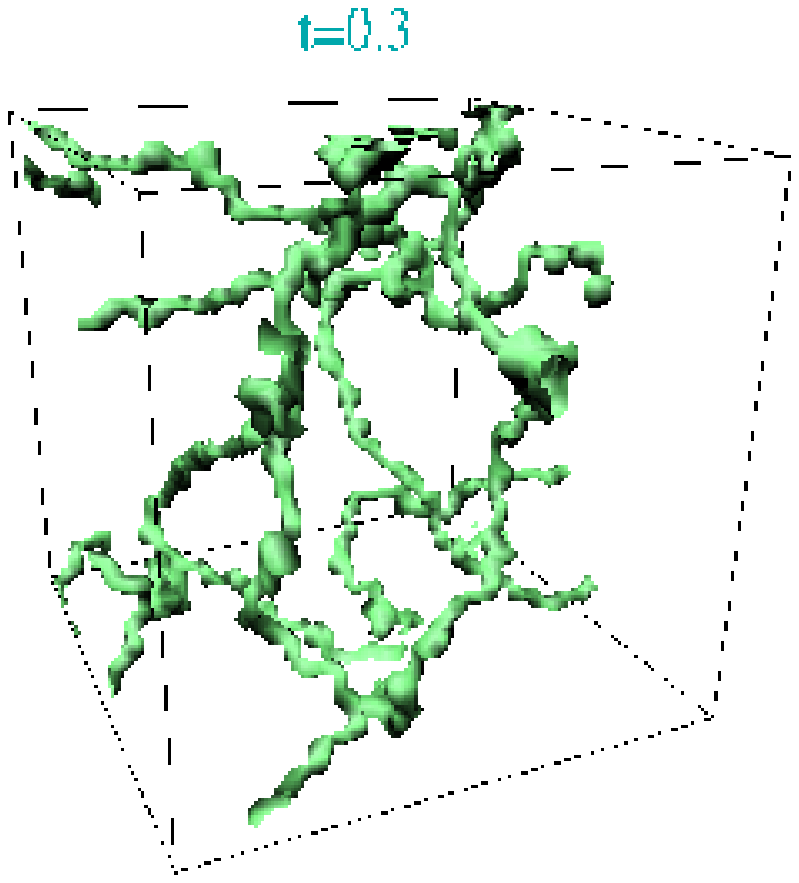,width=5cm}
\vspace{-.4cm}

\epsfig{file=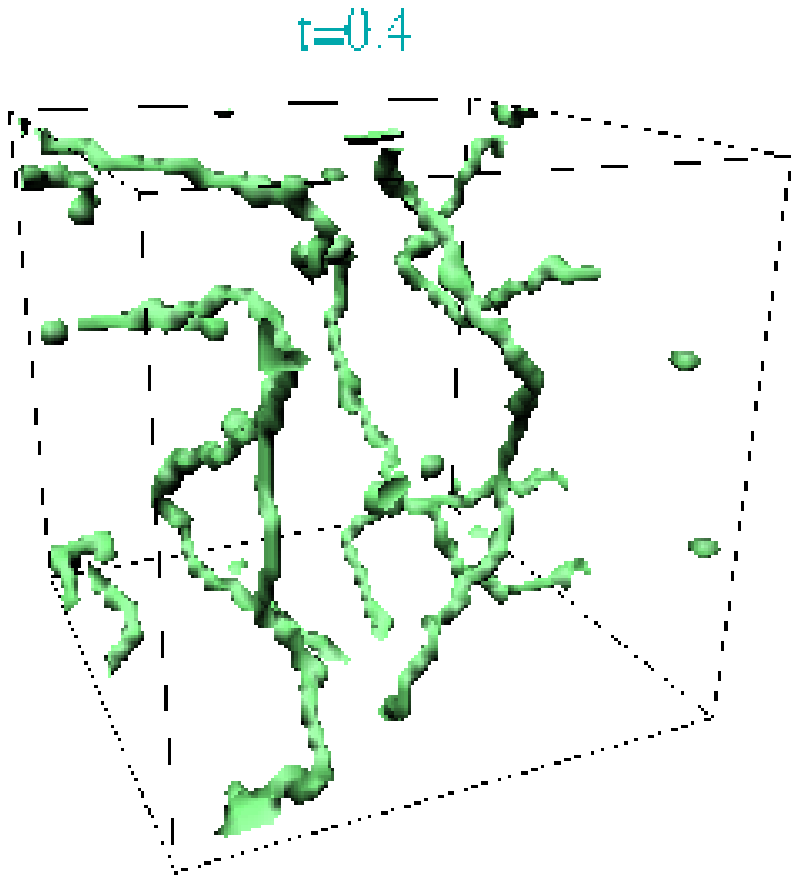,width=5cm}
\vspace{-1cm}
\caption{Time evolution of the winding number density during the
course of a first-order phase transition.}
\label{fig:kibble}
\end{figure}

As an example, we show in Fig.~\ref{fig:kibble} 
how the winding number density changes
during the phase transition. Initially, the system is full of fluctuating
strings, but after some time, some bubbles of the Higgs phase form and start
to expand. The bubbles collide and strings are trapped between them.
This is an example of the Kibble mechanism \cite{ref:kibble}.
With the technique presented here, 
we can measure the properties of the resulting
string networks and compare them with estimates based on 
non-gauged theories.

This approach can also be used to study various other 
properties of the phase transition.
One of the most interesting quantities is the bubble nucleation
rate. It can be measured in a straightforward way by taking a large number
of initial configurations from the supercooled Coulomb phase and evolving
each of them in time. If we plot the number of runs still remaining in
the symmetric phase after time $t$, we get an exponentially decreasing
curve, whose exponent gives the nucleation rate. This can be compared
with the analytical results \cite{ref:bubbles} 
to see, how well the analytical calculation
really works in gauge theories.

\section*{ACKNOWLEDGEMENTS}
I would like to thank M.~Hindmarsh for collaboration on this topic
and G.D.~Moore and K.~Rummukainen for discussions.
I acknowledge computing
support from the Sussex High Performance Computing
Initiative.

\end{document}